\documentclass[modern,longauthor]{aastex62}
\usepackage{amsmath}
\usepackage{comment}
\usepackage{xcolor}
\usepackage{hyperref}
\usepackage{anyfontsize}

\definecolor{reddish}{rgb}{0.813642,0.212345,0.196948}
\definecolor{blueish}{rgb}{0.0745098,0.52549,0.776471}
		
\hypersetup{
        colorlinks=true,        
        urlcolor=blueish,          
        pdfpagelabels=true,
        hypertexnames=true,
        plainpages=false,
        naturalnames=true,
        pdftitle={The Halo Drive},    
        pdfauthor={Kipping},     
        linkcolor=reddish,          
        citecolor=blueish,        
}

\newcommand{\pdf}{\mathrm{Pr}}
\newcommand{\taumax}{\tau_{\mathrm{max}}}

\shorttitle{Contact Inequality}
\shortauthors{Kipping et al.}

\begin{document}

\title{CONTACT INEQUALITY:\\
FIRST CONTACT WILL LIKELY BE WITH AN OLDER CIVILIZATION}

\correspondingauthor{David Kipping}
\email{dkipping@astro.columbia.edu}

\author[0000-0002-4365-7366]{{\fontsize{10.5}{12.6}\selectfont \textcolor{black}{David Kipping$^{1,2}$, Adam Frank$^{3}$ and Caleb Scharf$^{1}$}}}
\affil{$^1$Department of Astronomy,
Columbia University,
550 W 120th Street,
New York, NY 10027}
\affil{$^2$Center for Computational Astophysics,
Flatiron Institute,
162 5th Av.,
New York, NY 10010}
\affil{$^3$Department of Physics and Astronomy,
University of Rochester,
Rochester, NY 14627}




\begin{abstract}
First contact with another civilization, or simply another intelligence
of some kind, will likely be quite different depending on whether
that intelligence is more or less advanced than ourselves. If we
assume that the lifetime distribution of intelligences follows
an approximately exponential distribution, one might naively assume that the
pile-up of short-lived entities dominates any detection or contact scenario.
However, it is argued here that the
probability of contact is proportional to the age of said
intelligence (or possibly stronger), which introduces a selection effect. We
demonstrate that detected intelligences will have a mean age twice that
of the underlying (detected + undetected) population, using the exponential
model. We find that our first contact will most likely be with an older
intelligence, provided that the maximum allowed mean lifetime of the
intelligence population, $\taumax$, is $\geq e$ times larger than our own. 
Older intelligences may be rare but they disproportionality contribute to first 
contacts, introducing what we call a ``contact inequality'', analogous to
wealth inequality. This reasoning formalizes intuitional arguments and
highlights that first contact would likely be one-sided, with ramifications
for how we approach SETI.
\end{abstract}

\keywords{SETI --- technosignatures --- interstellar communication}

\section{Introduction}

The lifetime of a communicative civilization, $L$, plays a critical role in the
Drake Equation \citep{drake:1965,cirkovic:2004,maccone:2010,glade:2012}.
Little is known about the possible range
that this value can take \citep{burchell:2006}. Our limited temporal existence
provides a basis to estimate that $L$ likely typically takes a value greater
than or equal to modern civilization's age thus far. Pessimists might suggest
that the history of past human civilizations indicates that $L$ will be brief, no
greater than a few hundred years \citep{shermer:2002}. Optimists could equally
argue that we will soon pass a critical juncture where comparable civilizations
could ultimately enjoy long lifetimes, perhaps even billions of years
\citep{grinspoon:2004}.

Although Drake cast $L$ as the communicative lifetime, modern SETI has evolved
to include both deliberate and unintentional signatures of technology -
``technosignatures'' \citep{wright:2017}. We go further by relaxing the
assumption that the technosignature need originate from what we would recognize
as a ``civilization'' - the source is an intelligence of some kind 
(e.g. an artificial intelligence) which is
capable of producing detectable technological signatures. In what follows, we
consider $L$ as representing the lifetime over which techosignatures from this
intelligence manifest.

One basic question concerning this hypothetical intelligence is - what would
first contact look like? This has been the playground of science fiction
writers for generations, and clearly this question has existential consequences
for our way of life. Although we have no information about other intelligences
yet, it is not unreasonable to assume that the nature of this contact will
depend considerably upon the relative technological capabilities of this
newfound entity. Humanity would surely treat communication with a comparably
developed civilization in quite a different manner from one with far
greater technological capabilities. The longer lived an intelligence, $L$, the
greater the opportunity for technological development. Accordingly, the
probability distribution of the lifetime of detected intelligences will be
of central importance to our decisions regarding contact.

We note that it is of course possible for artifacts from an intelligence (or
indeed a civilization) to persist far longer than the age of that entity.
Sometimes referred to as artifact SETI, there is particular interest in
applying to this Solar System objects \citep{freitas:1983,lacki:2019,
wright:sol}, for example. However, the detection of an artifact from a now
extinct intelligence presents no opportunity for direct communication or
interaction (even if this is unclear from the initial detection).

In this work, we therefore ask - what is the likely age of a detected and
extant intelligence. Certainly speculation on this topic exists
elsewhere. Carl Sagan famously wrote that that civilizations were unlikely to
be in technological lockstep with us \citep{sagan:1994} and thus would either
be far less advanced or far more advanced. Since the less advanced ones would
be undetected, this simple argument suggests contact would be with an older
intelligence. Similarly, Stephen Hawking warned that contact would likely be
with a more advanced and thus potentially dangerous entity. In what follows, we
attempt to formalize the logic behind this problem and establish some
statistical results for $L$ using a simple but plausible analytic model.

\section{A Model for Technosignature Lifetime}

\subsection{Exponential Distribution for $L$}
\label{sub:exponential}

At its core, we are asking a statistical question - what is the
\textit{likely} age of a detected intelligence. The first requirement to make
progress is to assign a probability distribution for $L$. The simplest lifetime
model we can posit is an exponential distribution \citep{lawless:2011}. We do
not claim that this is necessarily the true distribution, and encourage the
reader to treat this as an approximate-yet-instructive model for
making analytic progress. Further discussion about the suitability of this
model is offered in the Discussion.

With such a model, amongst the ensemble of all intelligences that will ever
arise, there would be a large number of short-lived intelligences
(potentially such as ourselves) and a much smaller number of long-lived
counterparts. On this basis, one might naively posit that communication with
another intelligence would surely be with one of the more abundant short-lived
intelligences. We proceed by first writing down the probability density
function of $L$ given our exponential distribution assumption:

\begin{align}
\pdf(L|\tau) &= \tau^{-1} e^{-L/\tau},
\label{eqn:exponential}
\end{align}

where $\tau$ is the mean lifetime from the distribution. The exponential
distribution assumes that the so-called hazard function\footnote{The hazard
function is defined as the probability that an observed values lies between
$t$ and $t+\mathrm{d}t$, given that it is larger than $t$ for infinitesimal
$\mathrm{d}t$.} is constant over time - much like a decaying atomic nucleus.
Certainly more sophisticated lifetime formulae have been suggested for species
survival. For example, a Weibull distribution is a commonly used
generalization of the exponential, that enables a time-dependent (specifically
a power-law) hazard function \citep{lawless:2011}, but comes at the
expense of an extra unknown parameter.

We note that a power-law distribution has also been adopted in ecology
studies \citep{pigolotti:2005}, but we found it to exhibit several disadvantages
over the exponential. First, it does not have semi-infinite support and thus
requires truncation parameterized some additional bounding parameter, either
a minimum lifetime or a maximum. Since no clear minimum exists, bounding at
the maximum leads to a function which is only monotonically decreasing
for indices between 0 and 1. This leads to an overly restrictive distribution
compared to the exponential and for these reasons it is not used in what follows.

\begin{figure*}
\begin{center}
\includegraphics[width=15.0cm]{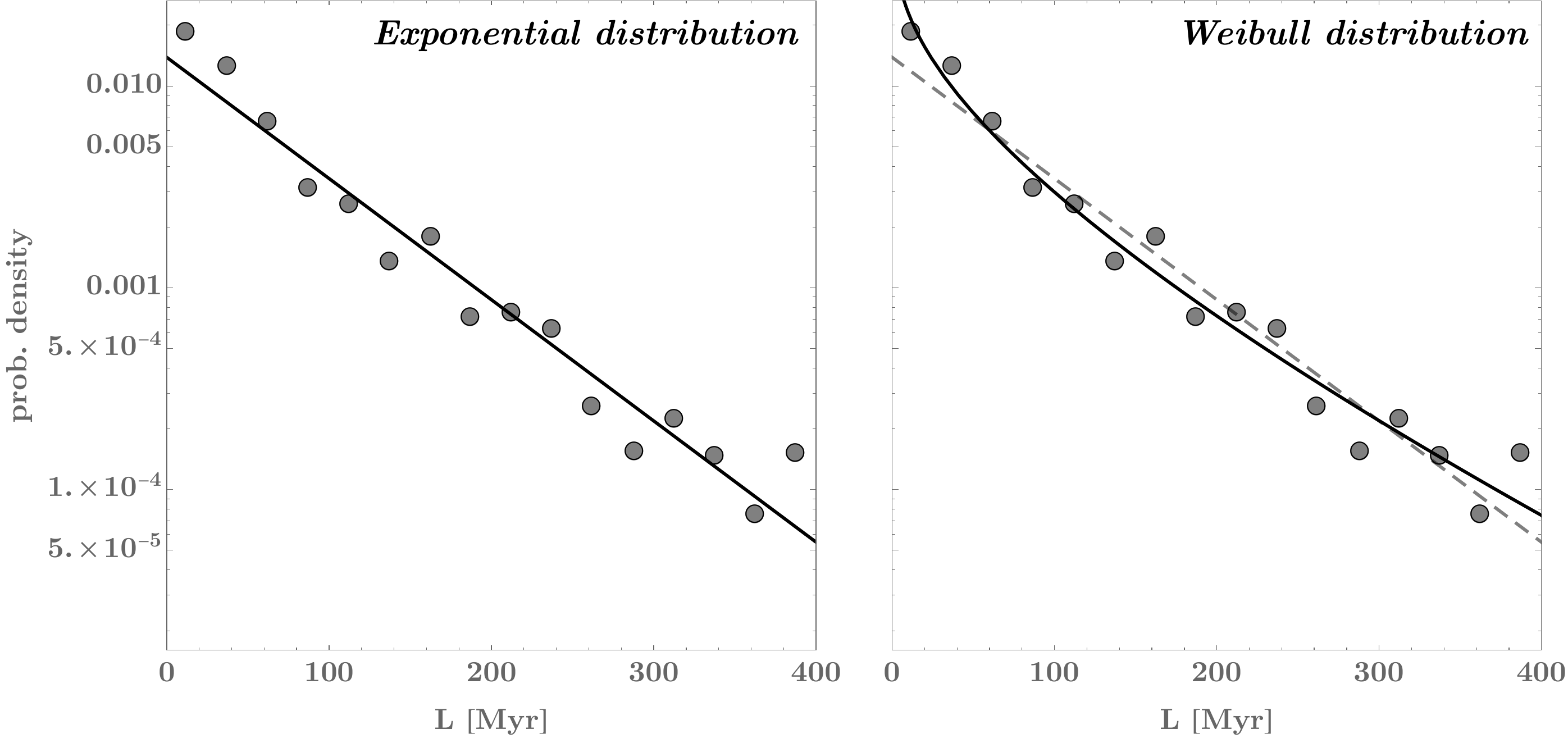}
\caption{
Probability distribution for the life span of families obtained from
\citet{benton:1993} using fossil evidence. On the left we fit an
exponential distribution to the data shown, whereas on the right we
show the more complicated Weibull distribution. Although we don't place
any emphasis on the specific parameters recovered, the data show that
an exponential distribution is a quite reasonable description of the
overall distribution, which compounded with it's simplicity makes it
attractive as a choice for modeling technosignature lifetimes.
}
\label{fig:fossil}
\end{center}
\end{figure*}

An exponential distribution, with its constant hazard function of $1/\tau$,
could be criticized as being unrealistic since a longer lived species
presumably has developed successful traits that improve its odds of future
survival \citep{shimada:2003}. On the other hand, as technology
advances, so too does an intelligence's capacity for self-destruction
\citep{cooper:2013}. If we consider the observed distribution for the
life span of families obtained from \citet{benton:1993} based on fossil
evidence (see Figure~\ref{fig:fossil}), the exponential distribution appears
quite capable of describing the overall pattern out to 400 million years. Of
course, intelligences producing technosignatures cannot be assumed to
necessarily follow the same distribution as these fossils, although this
give confidence that the model is at least plausible. Although
not a representative nor unbiased sample, we note that approximate
lifetimes of past human civilizations is also well-described by an
exponential distribution with $\tau=336$\,years\footnote{Using the
lifetimes reported at \href{http://energyskeptic.com/2019/part-2-how-long-do-civilizations-last-on-average-336-years/}{EnergySkeptic.com}}.

In the
absence of any other information, we invoke Ockham's razor in that
the simplest viable model is the presently favored one. Accordingly, we
will adopt the exponential distribution in what follows.

\subsection{Inferring the a-posteriori distribution of $\tau$}

Although we have a functional form for the probability distribution of $L$,
it is governed by a shape parameter, $\tau$ (mean lifetime), which one needs
to also assign. This would typically be handled through statistical inference.
For example, if we had $N$ known examples of intelligences with lifetimes
$\mathbf{L} = \{L_1,L_2,...,L_N\}^T$ (analogous to the data presented in
Figure~\ref{fig:fossil}), then we could write that the likelihood of measuring
these values for a mean lifetime equal to $\tau$ would be

\begin{align}
\pdf(\mathbf{L}|\tau) &= \prod_{i=1}^N \tau^{-1} e^{-L_i/\tau},
\end{align}

where one can see that the above is a straight-forward extension of
Equation~(\ref{eqn:exponential}). Conventionally, one would then
apply Bayes' theorem to constrain/measure $\tau$ using
$\pdf(\tau|\mathbf{L}) \propto \pdf(\mathbf{L}|\tau) \pdf(\tau)$.

Unfortunately, we do not have a sample of $L_i$ values, and thus our likelihood
function will certainly not be as constraining as this. Rather, we only know of
$N=1$ intelligence - ourselves. However, the problem is even worse than this
because we do not even know $L_1$ for this one datum. Human civilization has
been producing a technosignature for an age $A_{\oplus}$ years, and the
lifetime of this intelligence must at least exceed this value (i.e.
$L_{\oplus} \geq A_{\oplus}$). We emphasize that it's somewhat
unclear what numerical value to assign to $A_{\oplus}$ at this point. Although
we've been transmitting radio signals for ${\sim}10^2$\,years, one might argue
that an advanced civilization could remotely detect our settlements
\citep{kuhn:2015} and polluted atmosphere \citep{schneider:2010,lin:2014} as
unintentional technosignatures, which could increase $A_{\oplus}$.
Regardless, we will proceed symbolically for the moment.

The likelihood of observing one civilization with $L_1{>}A_{\oplus}$, given that
the mean lifetime is $\tau$, is given by

\begin{align}
\pdf(L_1{>}A_{\oplus}|\tau) &= \int_{A_{\oplus}}^{\infty} \pdf(L|\tau) \mathrm{d}L\nonumber\\
\qquad&= e^{-A_{\oplus}/\tau}.
\end{align}

In order to derive an \textit{a-posteriori} distribution for $\tau$,
conditioned upon the constraint that $L_1>A_{\oplus}$, we first need to write
down an \textit{a-priori} distribution for $\tau$. One is always free to choose
any prior one wishes, but a strongly informative prior, such as a tight
Gaussian, would naturally return a result which closely equals the prior.
In other words, one hasn't really learned anything and no inference really
occurred. Ideally, we wish to select a prior which is as uninformative as
possible \citep{jaynes:1968}. This is not simply a flat prior, since such
priors can place insufficient weight on small values, especially when the
parameter has high dynamic range. Instead, we can define an objective
Jeffrey's prior, which provides a means of expressing a scale-invariant
distribution via the Fisher information matrix,
$\mathcal{I}$ \citep{jeffreys:1946}:

\begin{align}
\pdf(\tau) \propto \sqrt{\mathrm{det}\mathcal{I}(\tau)}.
\end{align}

Evaluating the above, we obtain $\pdf(\tau) \propto \tau^{-1/2}$. Combining
the likelihood and prior together, we obtain

\begin{align}
\pdf(\tau|L_1{>}A_{\oplus}) &\propto \pdf(L_1{>}A_{\oplus}|\tau) \pdf(\tau),\nonumber\\
\qquad& \propto \tau^{-1/2} e^{-A_{\oplus}/\tau}.
\end{align}

To normalize the above, one must define an upper limit on $\tau$, for which
we use the symbol $\taumax$. At this point, it is also convenient to
work in temporal units of $A_{\oplus}$ in what follows, such that any
timescales used will always be in that unit. Accordingly, the posterior is

\begin{align}
\pdf(\tau|L_1{>}1) &= \frac{
\tau^{-1/2} e^{-1/\tau}
}{
2 \sqrt{\taumax} e^{-1/\taumax} - 2 \sqrt{\pi} \mathrm{erfc}[1/\sqrt{\taumax}]
}.
\label{eqn:taupost}
\end{align}

\begin{figure}
\begin{center}
\includegraphics[width=\columnwidth]{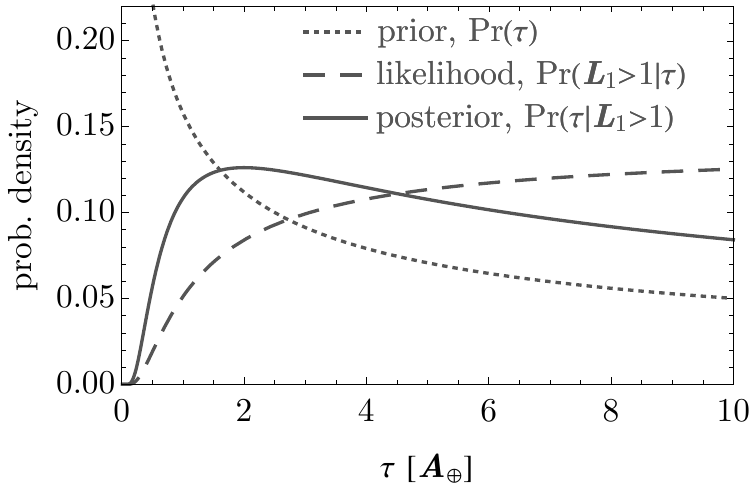}
\caption{
Comparison of the prior, likelihood and posterior distribution for
$\tau$ (the mean lifetime of intelligences producing technosignatures)
using $\taumax=10$\,Gyr, as an example. The mode of the posterior occurs at 2
as shown in the text.
}
\label{fig:taupost}
\end{center}
\end{figure}

We plot the posterior, with comparison to the prior and likelihood,
in Figure~\ref{fig:taupost}.

\subsection{Properties of the posterior}

There are several useful properties of the posterior above that we highlight.
First, Equation~(\ref{eqn:taupost}) has a maximum at $\hat{\tau} = 2$
(the mode), irrespective of $\taumax$, which can be demonstrated through
differentiation of the expression and setting to zero. If we set
$\tau=\hat{\tau}$, then the mean lifetime of an intelligence would be twice of
that of ourselves. But it's important to remember that this is the entire
lifetime of this intelligence, not it's age at the time of of their detection,
$A$. Assuming that the technosignature is no more or less likely to be detected
at any point during its manifested lifetime, then $A \sim \mathcal{U}[0,L]$
(where $\mathcal{U}$ denotes a uniform distribution). Accordingly, if
$\tau=\hat{\tau}$, then the mean age at the time of detection would $=1$ i.e.
our current age. Of course, fixing $\tau=\hat{\tau}$ does not correctly account
for the broad posterior distribution of $\tau$, but this exercise provides some
intuition as to why the modal value of $\tau$ occurs at 2.

Although the mode can be solved for independent of $\taumax$, it is somewhat
limited as an interpretable summary statistic. The \textit{expectation value} of
a distribution provides better intuition as to the ``typical'' value of the
distribution. This can be seen by simple consideration of the exponential
distribution. Its mode is zero but the average draw will be around the mean of
the distribution, not zero. We may calculate the \textit{a-posteriori}
expectation value for $\tau$ using

\begin{align}
E[\tau|L_1{>}1] &= \int_{\tau=0}^{\taumax} \tau \pdf(\tau|L{>}1)\,\mathrm{d}\tau\nonumber\\
\qquad&= \mu,
\end{align}

where we define the symbol

\begin{align}
\mu &\equiv \frac{1}{3} \Bigg( \frac{\taumax}{ 1 - \sqrt{\pi} \taumax^{-1/2} e^{1/\taumax}
\mathrm{erfc}[1/\sqrt{\taumax}]} - 2 \Bigg),
\end{align}

where $\mathrm{erfc}[x]$ is the complementary error function.
One may show that $\mu \simeq \taumax/3$ for $\taumax \gg 1$. The
dependency upon $\taumax$ can be understood by the fact that although the mode
of the distribution does not depend on $\taumax$, pushing the upper limit ever
higher naturally drags the tail out and thus pulls the expectation value over.

\subsection{Marginalized distribution for $L$}

Given that we have now obtained an \textit{a-posteriori} distribution for
$\tau$, we need to propagate that into a posterior distribution for $L$. As
discussed earlier, simply fixing $\tau$ to an \textit{a-posteriori} summary
statistic, like $\hat{\tau}$ or $\mathrm{E}[\tau|L_1{>}1]$, is inadequate, as it
does not propagate the (considerable) uncertainty on $\tau$ into the resulting
distribution. This propagation can be conducted through marginalizing out $\tau$
(i.e. integrating over $\tau$):

\begin{align}
\pdf(L|L_1{>}1) &= \int_{\tau=0}^{\taumax} \pdf(L|\tau) \pdf(\tau|L_1{>}1) \mathrm{d}\tau,\nonumber\\
\qquad&= \frac{
\sqrt{\pi} \mathrm{erfc}[\sqrt{L+1}/\sqrt{\taumax}]
}{
2 \sqrt{L+1} \big( \sqrt{\taumax} e^{-1/\taumax} - \sqrt{\pi} \mathrm{erfc}[1/\sqrt{\taumax}] \big)
}.
\end{align}

The above represents the probability distribution of the lifetime of
technosignature-producing intelligences, given the singular constraint imposed
by humanity's existence. It has a maximum at $L\to0$, which is a property shared
by the original exponential distribution used for $\pdf(L)$. We also note that
the expectation value satisfies $\mathrm{E}[L|L_1{>}1] = \mu$.

\section{Observationally Weighting the Model}

\subsection{Lifetime weighting}

The distribution $\pdf(L|L_1{>}1)$ describes the probability distribution of
the lifetime of intelligences producing detectable technosignatures. This is
the underlying true population - but it does not represent the intelligences
that we are most likely to detect. It's worth pausing to clearly distinguish
between detection and contact. If and when an intelligence is detected, that
detection may either be in the form of a directed attempt at communication on
their behalf, or it may simply be passive detection of their technology on our
behalf. Regardless, humanity's decision as to whether to send a message back -
to initiate contact - will be likely somewhat dependent on the technological
development and, by proxy, age ($A$) of said intelligence\footnote{Age
is subtlety distinct from lifetime and the difference between the two is
expounded upon more rigorously in the next subsection.}. If the
technosignature itself provides little information regarding the age, we would
be left with the \textit{a-priori} distribution - which is the focus of this
paper. Yet this distribution will not simply equal $P(A|L_1)$, since a critical
selection effect sculpts our observations that we will account for here.

The start time of these other intelligences is presumably arbitrary (except
when one pushes into timescales of $\gg$Gyr, over which time variability is
expected for the rates of star formation and high-energy astrophysical
phenomena e.g. SNe, AGNs, GRBs). A start time
10 million years ago is just as \textit{a-priori} likely as 100 years ago. Thus,
a longer lived intelligence is more likely to be detected than one which is very
short lived, since the requirement for contemporaneity (modulo the light cone)
is clearly sensitive to how long the technosignature persists. An equivalent
statement is that at any single snapshot in time (representing our current epoch
for example), the fraction of worlds that go on to produce long-lived
intelligences may be relatively rare, but their persistence through time means
that one must account for their overrepresentation amongst the extant
intelligences. This is simply a product of their longevity and is independent
of their activities or behavior. This situation is analogous to the ages of
trees in an old growth forest - if we assigned a unique identity to each tree
that will ever live, 1000+ year old trees are rare amongst the ensemble,
perhaps representing just 1\%, yet a
visit to the forest will show them to be seemingly more common due to their
longevity, for example comprising 10\% of the extant trees.

Accordingly, we will assume that the probability of detecting an intelligence's
technosignature is proportional to its lifetime, $L$. The validity of this
assumption is discussed later in the Discussion section, as well as an
explanation as to why distance does not affect the results presented hereafter.

This simple weighting will substantially change the picture, meaning that the
long tail of rare long-lived intelligences will have a considerable increase
on their relative probability of detection. We write that the probability
distribution of $L$, conditioned upon both a mean lifetime, $\tau$, and the
assumption of detection, $\mathcal{D}$, is

\begin{align}
\pdf(L|\tau,\mathcal{D}) \propto L \pdf(L|\tau),
\end{align}

or after normalization

\begin{align}
\pdf(L|\tau,\mathcal{D}) &= \frac{L}{\tau^2} e^{-L/\tau}.
\end{align}

Since we have already learnt $\tau$ from before, we can use this acquired
information to express a marginalized posterior for $L$ conditioned upon
both $\mathcal{D}$ and the fact $L_1>1$, using:

\begin{align}
\pdf(L|L_1{>}1,\mathcal{D}) &= \int_{\tau=0}^{\taumax} \pdf(L|\tau,\mathcal{D}) \pdf(\tau|L_1>1) \mathrm{d}\tau,\nonumber\\
\qquad&= W \Bigg(\frac{L e^{-L/\taumax}}{4 (L+1)^{3/2}}\Bigg) .
\end{align}

where

\begin{align}
W = \Bigg(\frac{2 \sqrt{\frac{L+1}{\taumax}} + \sqrt{\pi} e^{(L+1)/\taumax} \mathrm{erfc}[\sqrt{L+1}/\sqrt{\taumax}]
}{
\sqrt{\taumax} - \sqrt{\pi} e^{1/\taumax} \mathrm{erfc}[1/\sqrt{\taumax}]
}\Bigg)
\end{align}

We find that in the limit of $\taumax\gg1$, this distribution peaks at $2$. The
expectation value is given by

\begin{align}
E[L|L_1{>}1,\mathcal{D}] &= \int_{\tau=0}^{\infty} L \pdf(\tau|L{>}1,\mathcal{D})\,\mathrm{d}L,\nonumber\\
\qquad&= 2\mu.
\end{align}

\begin{figure}
\begin{center}
\includegraphics[width=\columnwidth]{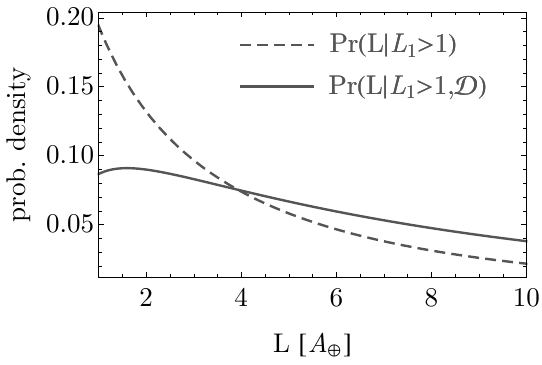}
\caption{
Comparison of the marginalized posterior probability for the age of
intelligences, $L$, for the ensemble population ($\pdf(L|L_1{>}1)$) and
the detected population ($\pdf(L|L_1{>}1,\mathcal{D})$). Here we 
adopt $\taumax=10$.
}
\label{fig:Lpost}
\end{center}
\end{figure}

For comparison, without the conditional $\mathcal{D}$, the
\textit{a-posteriori} expectation value was $\mu$ but including it doubles it.
We plot the posterior $\pdf(L|L_1{>}1,\mathcal{D})$, and compare it to
$\pdf(L|L_1{>}1)$, in Figure~\ref{fig:Lpost}.

\subsection{The age distribution of detected intelligences}
\label{sub:age}

The final step is to account for the fact that detection would not occur with
an intelligence at the end of its lifetime, $L$, but rather one drawn randomly
from across its lifespan. In other words, an intelligence's age (at the time of
detection) does not equal its lifetime. If we assume that the age at the time
of detection is uniformly distributed from 0 to $L$, then

\begin{align}
\pdf(A|L_1{>}1,\mathcal{D}) &= \int_{\tau=0}^{\taumax} \mathcal{U}[0,L] \pdf(L|L_1{>}1,\mathcal{D}) \mathrm{d}L,\nonumber\\
\qquad&= \frac{ \sqrt{\pi} \mathrm{erfc}[\sqrt{A+1}/\sqrt{\taumax}] }{
2 \sqrt{A+1} \big( \sqrt{\taumax} e^{-1/\taumax} - \sqrt{\pi}\mathrm{erfc}[1/\sqrt{\taumax}] \big).
}
\end{align}

Equipped with our final form for the \textit{a-posteriori} probability
distribution of the age of detected intelligences, we can deduce several
basic properties. First, it is interesting to ask whether the civilization
is likely to be older or younger than our own. The probability than the
civilization is older is given by

\begin{align}
\pdf(A{>}1|L_1{>}1,\mathcal{D}) =& \int_{A=1}^{\infty} \pdf(A|L_1{>}1,\mathcal{D}) \mathrm{d}A,\nonumber\\
\qquad=& \Big(\sqrt{\taumax} e^{-1/\taumax} \nonumber\\
\qquad& - \sqrt{2\pi}e^{1/\taumax}\mathrm{erfc}[\sqrt{2}/\sqrt{\taumax}]\Big)\nonumber\\
\qquad& /\Big(\sqrt{\taumax} - \sqrt{\pi}e^{1/\taumax}\mathrm{erfc}[1/\sqrt{\taumax}]\Big).
\end{align}

A useful summary statistic to interpret the above is the median -
above which half the cases will lie. This may be solved for by setting
the above to 0.5 and numerically solve for $\taumax$,
which gives the result that if $\taumax>2.6776...\simeq e$, then the age
of a detected intelligence will most likely exceed that of our own i.e.
$\pdf(A{>}1|L_1{>}1,\mathcal{D}) > 0.5$. In other words, if this condition
is true than we are most likely to detect an older intelligence than
ourselves.

It's important to remember that $\taumax$ does not represent the maximum
allowed lifetime of a civilization, $L$ - rather it's simply the maximum
\textit{a-priori} mean lifetime. Fundamentally, there is no obvious reason
why $\taumax$ could not be many billions of years \citep{grinspoon:2004},
and thus detection would almost always occur with an older civilization,
however one defines $A_{\oplus}$.

The expectation value for the intelligence's \textit{age} is given by simply
$\mu$, whereas we found the expectation value for their \textit{lifetime} to
be $2\mu$. Since $\mathrm{E}[A|L_1{>}1] = \mathrm{E}[L|L_1{>}1]/2$,
then we can see that the effect of including this observational bias is
that the mean age of detected, and thus contacted, intelligences is twice
that of the overall population - as expected.

\subsection{Contact inequality}

Using our results, it is instructive to compare the underlying age
population, $\pdf(A|L_1{>}1)$, with the population which goes on to
be detected, $\pdf(A|L_1{>}1,\mathcal{D})$. Recall that
$A$ is age of the intelligence at the time of detection/contact,
whereas $L$ is the total lifetime of said intelligence. The fact that
older intelligences are assumed in this work to be more likely to be
detected, and thus contacted (by virtue of having simply more
opportunities to do so), introduces an inequality. The rare
long-lived intelligences make a disproportionate number of contacts.

This ``contact inequality'' can be thought of as being analogous to
wealth inequality in economics. One way to quantify the degree of
inequality comes from the Gini coefficient \citep{gini:1909},
which takes the value of 1 for a maximally unequal distribution,
and 0 for a fully equal one. It may be calculated for a probability
density function $\pdf(x)$ using

\begin{align}
G &= \frac{1}{2\mu} \int_{0}^{\infty} \int_{0}^{\infty} \pdf(x) \pdf(y) |x-y|\,\mathrm{d}x\,\mathrm{d}y.
\end{align}

Although we were not able find a closed-form solution to the
above using $\pdf(x)=\pdf(A|L_1>1,\mathcal{D})$, one may numerically
integrate the expression for a specific choice of $\taumax$.

We argue here that a conservative choice of $\taumax$ is one which
causes our current age to be the median age of the entire population
of technosignature producing intelligences. This is a form of the
mediocrity principle, since we posit humanity lives close to the
center of the age-ordered list of intelligences in the cosmos
\citep{gott:1993,simpson:2016}. It requires us to solve $\taumax$
such that

\begin{align}
\int_{A=0}^{1} \pdf(A|L_1>1) &= \frac{1}{2}.
\end{align}

We solved the above numerically and obtain $\taumax=9.43$. This
also somewhat passes the astronomer's logic of going up by an
order-of-magnitude as one's upper limit on a variable. However,
we suggest here that this limit is somewhat conservative though,
since it makes million-/billion-year intelligences essentially
non-existent, which is itself a strong assumption.

Nevertheless, using $\taumax=9.43$, we compute a Gini coefficient
of 0.57. The value doesn't grossly change by varying $\taumax$.
For example, setting $\taumax=10^3$ increases $G$ to 0.63,
and decreasing it to $\taumax=1$ yields $G=0.52$. Interestingly,
we find that in the limit of $\taumax \to 0$ (which would make
humanity an incredibly long lived civilization)\footnote{We also
note that $\lim_{\taumax\to\infty}G=1$.}, $G\to0.5$.
Thus, under the assumptions of our simple model, we find that
$G\geq0.5$, which is similar to the wealth inequality of many
developed nations.

\begin{figure*}
\begin{center}
\includegraphics[width=15.0cm]{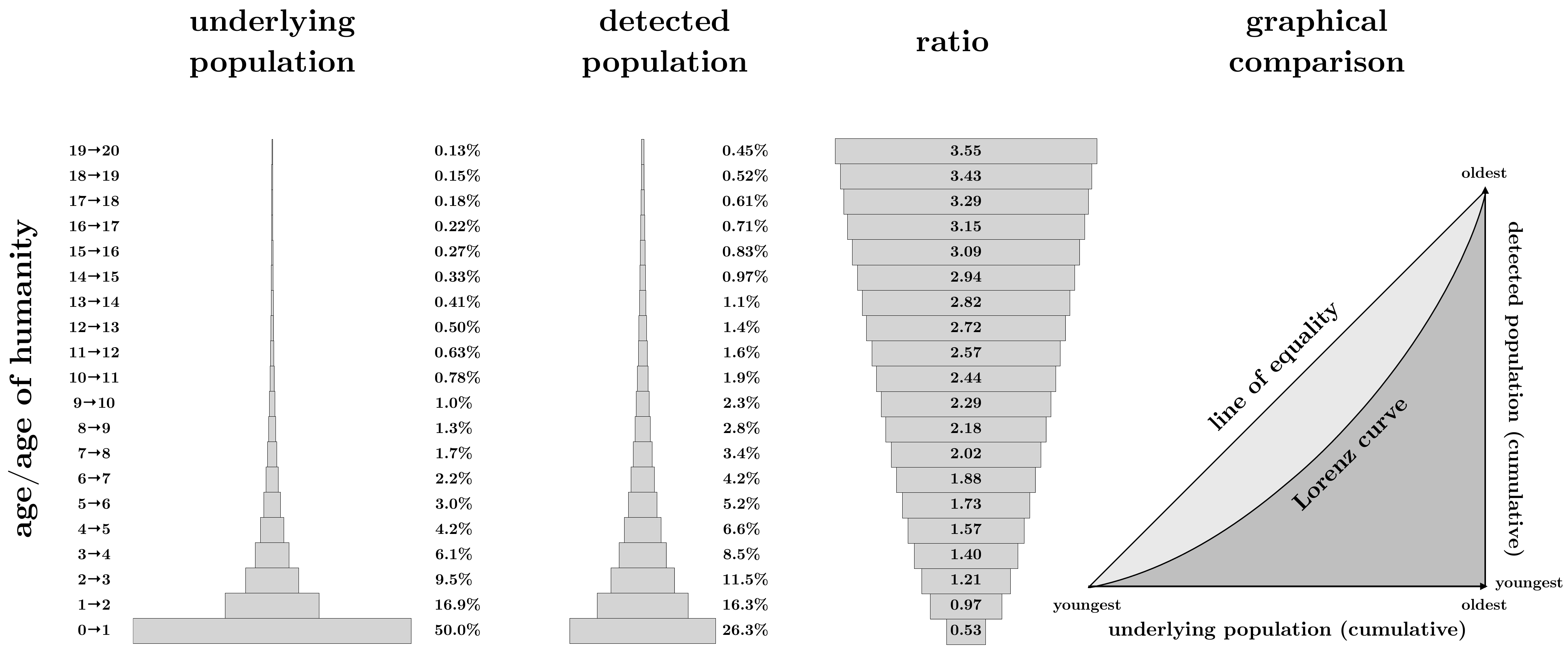}
\caption{
Using $\taumax=9.43$, one can set the \textit{a-posteriori}
distribution of intelligence ages such that humanity lives
at the median (far left). The exponential distribution
assumed heavily weights the population towards younger
civilizations, most of which will not progress into older
ones. However, older intelligences have more opportunities
to contact others, simply by their greater age, which skews
the distribution of the contacted population (mid-left).
Taking the ratio of the two (mid-right), the ``contact
inequality'' is apparent - which can also be visualized as
a Lorenz curve (far-right).
}
\label{fig:demographics}
\end{center}
\end{figure*}

To visualize the inequality, we show a stacked histogram of
the \textit{a-posteriori} age distribution of intelligences
in Figure~\ref{fig:demographics} using $\taumax=9.43$.
Specifically, one can see the effect of the bias weighting longer
lived intelligences. We find that the top 1\% of the oldest
intelligences are over-represented in the fraction of first
contacts by a factor of 4.

\section{Discussion}
\label{sec:discussion}

In this work, we have suggested a simple model for the lifetime distribution of
civilizations (or more generally intelligences) producing technosignatures -
specifically an exponential distribution. This is motivated by its monotonic,
single-parameter form and is a simple but effective description of the
lifetime of biological families on Earth. Amongst these hypothetical
intelligences, we may plausibly detect their technosignatures in the coming
years, which may either take the form of direct contact or open the door for us
to contact them. We have argued that the fact that longer lived intelligences simply
have had more time available to them makes them more likely to be detected -
and thus the contacted population is weighted towards older intelligences.
We highlight that a similar conclusion was independently arrived at by
\citet{balbi:2018}.

Another framing of the above is that at any given time, the number of extant
long-lived intelligences is disproportionality represented simply by the fact
they persist longer than their short-lived counterparts.

We are able to establish that the expectation age of a contacted intelligence
is twice that of the ensemble, without any assumption about the maximum mean
lifespan of this population. Further, we show that if the maximum mean
lifespan of intelligences is any greater than ${\sim}e$ times our current age,
then we will most likely detect an older intelligence than ourselves.

Finally, we use this simple model to show that a ``contact inequality'' should
exist, where the older intelligences represent a disproportionate fraction of
galactic first contacts. Using this analogy, we can define a Gini coefficient
to quantify the inequality, which we show must be greater than $0.5$ for
any choice of the maximum mean intelligence lifetime.

In this discussion, we would like to highlight two points. First, in what ways
might this model be invalid? And second, what are the consequences for us if
this model is correct?

\subsection{Validity of the employed model}

First, we fully acknowledge here that the exponential distribution model is
indeed extremely simplistic and may not fully describe the true
distribution. The hazard function is a constant with respect to age and
it's deeply unclear whether a more advanced intelligence poses a greater
risk to itself through emerging technologies (e.g. \citealt{cooper:2013}),
or, on the other hand, is more likely to persist due to their track record
of survival thus far. The lifespan of biological families from fossil
evidence shows that an exponential distribution may not always be the
best fit, but it does broadly capture the overall behavior (see
\citealt{shimada:2003} and Figure~\ref{fig:fossil}). It also satisfies the
basic expectation of a monotonically decreasing smooth function. Without
any other evidence in hand, we argue that at present there is no justification
for invoking a more complex model.

The assumption of lifetime-weighted contact also deserves scrutiny.
In this work, we have very simply assumed that the longer an intelligence
lasts, the more opportunities it has to be spotted. For example, if a
civilization builds a beacon which lasts for an interval $L$ at some
random point in the Universe's history, the probability that we will
detect that beacon must be directly proportional to $L$. But of course
one could challenge this picture from both the direction of increased
or decreased detectability\footnote{Moreover, it would be interesting
to consider non-monontonic models with functional dependencies of
detectability versus lifetime (but not to be confused with age).}.

For example, as an intelligence becomes more advanced, it could construct
more powerful beacons, with greater range, at lower cost, and in greater
number \citep{benford:2008}, even sending them out between the stars to
add coverage. Those are intentional contact scenarios, but even unintended
technosignatures might be argued to become more detectable as intelligences
advance, such as the production of Dysonian artifacts \citep{dyson:1960}.
On this basis, one might conclude that our assumption here that the
probability of contact is proportional to $L$ greatly underestimates the true
value. If so, then older intelligences would dominate the number of first
contacts by an even more extreme degree, raising the Gini index yet
higher\footnote{We did attempt to repeat our study assuming a proportionality
of $L^n$, where $n$ is a free-index, but were not able to make analytic
progress.}. This fundamentally does not change our hypothesis that a contact
inequality likely exists, in fact it exacerbates the inequality.

On the other hand, one might argue that as intelligences develop, their
detectability decreases. Science fiction writer Karl Schroeder captures this
hypothesis in his twist on Arthur C. Clarke's famous line ``Any sufficiently
advanced civilization is indistinguishable from nature'' \citep{schroeder}.
They might also simply lose interest in communicating with far less advanced
intelligences and elect to hide themselves \citep{smart:2012,cloaking}.
If their detectable presence is suddenly eliminated altogether, then they are
technically no longer a member of the assumed underlying population - which
is specifically one which produces (potentially detectable) technosignatures.
Thus, they are effectively extinct and thus don't actually affect the arguments
laid out here. However, if the detectability of intelligences diminishes with
age, in particular in a way such that the time-integrated probability of
detection culminates in a scaling of $L^{\alpha}$ where $\alpha<0$,
then this would
reverse our conclusion - contact would likely occur with less advanced members
of the population.

Although we certainly don't discount this possibility, extrapolation of our
own behavior does not generally favor this conclusion. Whilst radio leakage
into space has been decreasing, many other aspects of human's detectability
projected into the future suggest that we could still be easily found
through other technosignatures. Some examples include space mining
\citep{forgan:2011}, leakage from relativistic light sails
\citep{guillochon:2015}, thermal heat islands, \citep{kuhn:2015}, our polluted
atmosphere \citep{schneider:2010,lin:2014}, geostationery satellites
\citep{socas:2018}, geoengineering projects \citep{gaidos:2017}, photovoltaic
cells \citep{lingham:2017}, space
weathering monitoring systems \citep{kipping:2019} and ever growing energy
needs \citep{wright:2014}. We thus consider that if our own experience and
future projections are in any way representative, a future decrease in our
technosignature detectability would likely require a deliberate and expensive
effort, which is itself unlikely to be considered a good use of resources in
the absence of any evidence for other intelligences. Accordingly, we argue
that such a scenario is unlikely to dominate until intelligences become much
older than our own - which is essentially captured by our assumption that
$\taumax$ is an order-of-magnitude greater than our current age.

Together, whilst we accept that our model is surely an oversimplification,
the qualitative result that older intelligences should be overrepresented
in the ensemble of detections may actually be quite robust.

\subsection{A note on distance}

Our detection bias model assumes that the probability of detection is
proportional to an intelligence's lifetime, but the distance to that
intelligence does not feature. Why not? Certainly, closer intelligences
will be more likely to be detected than more distant ones, since signals
generally decrease as $1/d^2$. But this work only concerns itself
with the lifetime distribution of detected intelligences, not their
distance (which can be thought of as being marginalized over). The real
question for this work is - do we expect there to be some off-diagonal
covariance between lifetime and distance of the detected population? More
simply, is there any reason to suspect that the intrinsic lifetimes of detected
intelligences is dependent upon their displaced location from the Earth?

As discussed in detail in the last section, one could invoke an
argument that longer lived intelligences are more detectable, which
would exacerbate the contact inequality result of this work. Only if
detectability rapidly diminished with time would our basic
conclusion change.

A separate aspect to the distance issue, is not with detectability per
say, but rather with intrinsic lifetimes varying with distance. Do
we expect an intelligence's lifetime to depend upon how far away from us
they are? At distances of hundreds, even thousands, of light years - the
answer is no. There is
nothing inherently special about where we live and thus a civilization
emerging a few hundred light years should not have any particular
reason to live longer or shorter than ourselves. Extending further
afield, where effects such as galactic chemical gradients
\citep{gonzales:2001}, supernovae rates \citep{lineweaver:2004},
active galactic nuclei \citep{balbi:2017,lingam:2019},
stellar encounter rates \citep{mctier:2020} may vary, would indeed
require formally building a model which described this covariance.
Accordingly, the results of our work should be understood to be
formally only applicable to where cases where $L$ is not expected
to be intrinsically linked to location, such as our local stellar
neighborhood.

\subsection{Implications}

Let's proceed under the assumption that the hypothesis is correct:
probabilistically, we are more likely to make first contact with an
intelligence that is considerably older than ourselves. It should be
noted that this age difference could be quite extreme, perhaps
millions or even billions of years, in principle. Although age does
not necessarily ensure greater technological advancement, that is the
obvious expectation from such a scenario. Of course, we may never detect
any technosignatures and thus never have the opportunity for first contact,
but under the premise that we will one day succeed, it is interesting to
ask what the implications of our suggested contacted inequality are.

Some have voiced concerns that humanity's historical record of encounters
between societies of different technological capabilities generally ends poorly
for the less advanced entity. Of course, it's unclear that human behavior
can be extrapolated to another intelligence that is far older than ourselves.
Accordingly, we prefer to avoid speculating about the impact of such a contact
directly.

However, the contact inequality hypothesis does have significant bearing on
our own active searches for technosignatures. Focusing on searching for
technology similar to that of our own may be unlikely to lead to success.
If an intelligence is much more advanced than us, then planet-integrated
transient signatures associated with disequilibrium
(such as climate change and pollution) are less likely to be the means of
detection, since they are simply unsustainable for a long-lived entity.

Further, to ensure their own survival, such intelligences may have relocated
or expanded their presence off-world, thus favoring technosignatures associated
with such activities. Ultimately, this work concerns itself with a formalism
for establishing the hypothesis rather the consequences of it. But from our
work, we encourage the formalism and prediction established here to be
considered in future efforts to seek out technosignatures, including more
detailed exploration of the assumptions and analytic forms of civilization
longevity and technological age.

\newpage
\section*{Acknowledgements}

DK is supported by the Alfred P. Sloan Foundation.
CS acknowledges support from the NASA Astrobiology Program through
participation in the Nexus for Exoplanet System Science and NASA Grant NNX15AK95G.
Special thanks to Tom Widdowson, Mark Sloan, Laura Sanborn, Douglas Daughaday, Andrew Jones, Jason Allen, Marc Lijoi, Elena West, Tristan Zajonc, Chuck Wolfred, Lasse Skov, Martin Kroebel \& Geoff Suter.




\begin{thebibliography}{}
\bibitem[\protect\citeauthoryear{Balbi \& Tombesi}{2017}]{balbi:2017} 
Balbi, A. \& Tombesi, F., ``The habitability of the Milky Way during the
active phase of its central supermassive black hole'', 2017, Sci. Rep., 7, 16626
\bibitem[\protect\citeauthoryear{Balbi}{2018}]{balbi:2018} 
Balbi, A., ``The Impact of the Temporal Distribution of Communicating Civilizations on Their Detectability'', 2018, Astrobiology, 18, 54
\bibitem[\protect\citeauthoryear{Benford et al.}{2008}]{benford:2008} 
Benford, G., Benford, J., Benford, D., 2008, ``Searching for Cost Optimized Interstellar
Beacons'', arXiv e-prints:0810.3966
\bibitem[\protect\citeauthoryear{Benton}{1993}]{benton:1993} 
Benton, M.~J., 1993, ``The Fossil Record'', Vol. 2 (Chapman \& Hall, London, 1993)
\bibitem[\protect\citeauthoryear{Burchell}{2006}]{burchell:2006} 
Burchell, M.~J., 2006, ``W(h)ither the Drake equation'',  Int. J. Astrobiology, 5, 243
\bibitem[\protect\citeauthoryear{\'Cirkovi\'c}{2004}]{cirkovic:2004} 
\'Cirkovi\'c, M.~M., 2004, ``The Temporal Aspect of the Drake Equation and SETI'',
Astrobiology, 4, 225
\bibitem[\protect\citeauthoryear{Cooper}{2013}]{cooper:2013} 
Cooper, J., 2013, ``Bioterrorism and the Fermi Paradox'', Int. J. Astrobiology, 12, 144
\bibitem[\protect\citeauthoryear{Drake}{1965}]{drake:1965} 
Drake, F.~D, 1965, ``The Radio Search for Intelligent Extraterrestrial Life'',
eds. Mamikunian G, Briggs MH. pp. 323–345.
\bibitem[\protect\citeauthoryear{Dyson}{1960}]{dyson:1960} 
Dyson, F.~J., 1960, ``Search for Artificial Stellar Sources of Infrared Radiation'',
Science, 131, 1667
\bibitem[\protect\citeauthoryear{Forgan \& Elvis}{2011}]{forgan:2011} 
Forgan, D.~H. \& Elvis, M., 2011, ``Extrasolar Asteroid Mining as Forensic Evidence for
Extraterrestrial Intelligence'', Int. J. Astrobiology, 10, 307
\bibitem[\protect\citeauthoryear{Freitas}{1983}]{freitas:1983} 
Freitas. R.~A.~Jr., ``The search for extraterrestrial artifacts (SETA)'', JBIS, 36, 501
\bibitem[\protect\citeauthoryear{Gaidos}{2017}]{gaidos:2017} 
Gaidos, E., 2017, ``Transit detection of a starshade at the inner lagrange point
of an exoplanet'', MNRAS, 469, 4455
\bibitem[\protect\citeauthoryear{Gini}{1909}]{gini:1909} 
Gini, C., 1909, ``Concentration and dependency ratios'' (in Italian).
English translation in Rivista di Politica Economica, 87 (1997), 769–789.
\bibitem[\protect\citeauthoryear{Glade et al.}{2012}]{glade:2012} 
Glade, N., Ballet, P., Bastien, O., 2012, ``A stochastic process approach of the drake equation parameters'', Int. J. Astrobiology, 11, 103
\bibitem[\protect\citeauthoryear{Gonzalez}{2001}]{gonzales:2001}
Gonzalez, G., Brownlee, D., Ward, P., 2001, ``The Galactic Habitable Zone: Galactic
Chemical Evolution'', Icarus, 152, 185.
\bibitem[\protect\citeauthoryear{Gott}{1993}]{gott:1993} 
Gott, J.~R. III., 1993, ``Implications of the Copernican principle for our future
prospects'', Nature, 363, 315
\bibitem[\protect\citeauthoryear{Grinspoon}{2004}]{grinspoon:2004} 
Grinspoon, D., 2004, ``Lonely Planets'', Ecco Press, New York
\bibitem[\protect\citeauthoryear{Guillochon \& Loeb}{2015}]{guillochon:2015} 
Guillochon, J. \& Loeb, A., 2015, ``SETI Via Leakage From Light Sails in Exoplanetary
Systems'' ApJ, 811, 20
\bibitem[\protect\citeauthoryear{Jaynes}{1968}]{jaynes:1968} 
Jaynes, E.~T., 1968, ``Prior probabilities'', IEEE Transactions on Systems
Science and Cybernetics, 4, 227
\bibitem[\protect\citeauthoryear{Jeffreys}{1946}]{jeffreys:1946} 
Jeffreys, H., 1946, ``An invariant form for the prior probability in estimation
problems'', Proc. Royal Society of London A., Mathematical and Physical Sciences, 186, 453
\bibitem[\protect\citeauthoryear{Kipping \& Teachey}{2016}]{cloaking} 
Kipping, D.~M. \& Teachey, A., 2016, ``A Cloaking Device for Transiting Planets'',
MNRAS, 459, 1233
\bibitem[\protect\citeauthoryear{Kipping}{2019}]{kipping:2019} 
Kipping, D.~M., 2019, ``Transiting Quasites as a Possible Technosignature'', RNAAS, 3, 91
\bibitem[\protect\citeauthoryear{Kuhn \& Berdyugina}{2015}]{kuhn:2015} 
Kuhn, J.~R. \& Berdyugina, S.~V., 2015, ``Global warming as a detectable thermodynamic marker of Earth-like extrasolar civilizations: the case for a telescope like Colossus'', Int. J. Astrobiology, 14, 401
\bibitem[\protect\citeauthoryear{Lacki}{2019}]{lacki:2019} 
Lacki, B., 2019, PASP, ``A Shiny New Method for SETI: Specular Reflections from Interplanetary Artifacts'', 131, 084401
\bibitem[\protect\citeauthoryear{Lawless}{2011}]{lawless:2011} 
Lawless, J.~F., 2011, ``Statistical models and methods for lifetime data'', vol. 362. Wiley, New York
\bibitem[\protect\citeauthoryear{Lin et al.}{2014}]{lin:2014} 
Lin, H.~W, Abad, G.~G. \& Loeb, A., 2014, ``Detecting Industrial Pollution in the Atmospheres of Earth-like Exoplanets'', ApJL, 792, L7
\bibitem[\protect\citeauthoryear{Lineweaver et al.}{2004}]{lineweaver:2004}
Lineweaver, C.~H., Fenner, Y., Gibson, B.~K., 2004, ``The Galactic Habitable Zone and the Age
Distribution of Complex Life in the Milky Way'', Science, 303, 59
\bibitem[\protect\citeauthoryear{Lingam \& Loeb}{2017}]{lingham:2017} 
Lingam, M. \& Loeb, A., 2017, ``Natural and artificial spectral edges in exoplanets'',
MNRAS, 470, 82
\bibitem[\protect\citeauthoryear{Lingam et al.}{2019}]{lingam:2019} 
Lingam, M., Ginsburg, I., Bialy, S., 2019, ``Active Galactic Nuclei: Boon or Bane for Biota?'',
ApJ, 877, 62
\bibitem[\protect\citeauthoryear{Maccone}{2010}]{maccone:2010} 
Maccone, C., 2010, ``The Statistical Drake Equation'',
Acta Astronautica, 67, 1366
\bibitem[\protect\citeauthoryear{McTier et al.}{2020}]{mctier:2020}
McTier, M., Kipping, D., Johnston, K., 2020, ``8 in 10 Stars in the Milky Way Bulge experience stellar encounters within 1000 AU in a gigayear'', MNRAS, 495, 2105
\bibitem[\protect\citeauthoryear{Pigolotti et al.}{2005}]{pigolotti:2005} 
Pigolotti, S., Flammini, A., Marsili, M., Maritan, A., 2005, ``Species lifetime distribution for simple models of ecologies'', PNAS, 102, 15747
\bibitem[\protect\citeauthoryear{Sagan}{1994}]{sagan:1994} 
Sagan, C., 1994, ``Pale Blue Dot: A Vision of the Human Future in Space'',
Random House, New York NY, pp. 352-354
\bibitem[\protect\citeauthoryear{Schneider et al.}{2010}]{schneider:2010} 
Schneider, J., L\'eger, A., Fridlund, M., White, et al., 2010, ``The Far Future of Exoplanet Direct Characterization'', Astrobiology, 10, 121
\bibitem[\protect\citeauthoryear{Schroeder}{2003}]{schroeder} 
Schroeder, K., 2011, ``The Deepening Paradox'', blog post, www.kschroeder.com/weblog/the-deepening-paradox
\bibitem[\protect\citeauthoryear{Shermer}{2002}]{shermer:2002} 
Shermer, M., 2002, ``Why ET Hasn't Called'', Scientific American, 287, 33
\bibitem[\protect\citeauthoryear{Shimada et al.}{2003}]{shimada:2003} 
Shimada, T., Yukawa, S., Ito, N., 2003, ``Life-Span of Families in Fossil Data Forms q-Exponential Distribution'', Int. J. Mod. Phys. C, 14, 1267
\bibitem[\protect\citeauthoryear{Simpson}{2016}]{simpson:2016} 
Simpson, F., 2016, ``Apocalypse Now? Reviving the Doomsday Argument'', arXiv e-prints:1611.03072
\bibitem[\protect\citeauthoryear{Smart}{2012}]{smart:2012} 
Smart, J.~M., ``The transcension hypothesis: Sufficiently advanced civilizations
invariably leave our universe, and implications for METI and SETI'', Acta Astronautica, 78, 55
\bibitem[\protect\citeauthoryear{Socas-Navarro}{2018}]{socas:2018} 
Socas-Navarro, H., 2018, ``Possible Photometric Signatures of Moderately Advanced Civilizations: The Clarke Exobelt'', ApJ, 855, 110
\bibitem[\protect\citeauthoryear{Wright}{2014}]{wright:2014} 
Wright, J.~T., Mullan, B., Sigurdsson, S., Povich, M. S., 2014, ``The G-hat Infrared Search for Extraterrestrial Civilizations with Large Energy Supplies. I. Background and Justification'', ApJ, 792, 26
\bibitem[\protect\citeauthoryear{Wright}{2018}]{wright:sol} 
Wright, J.~T., 2018, ``Prior indigenous technological species'', Int. J. Astrobiology, 17, 96
\bibitem[\protect\citeauthoryear{Wright}{2017}]{wright:2017} 
Wright, J.~T., 2017, ``Exoplanets and SETI'', Handbook of Exoplanets, Springer International Publishing AG,
part of Springer Nature, 2018, id.186
\end{thebibliography}
\end{document}